\documentclass[
preprint,
footinbib,
amsmath,amssymb,
aps,
pra,
]{revtex4-2}
\usepackage[margin=1in]{geometry}
\usepackage[utf8]{inputenc}
\usepackage[T1]{fontenc}
\usepackage{amsmath}
\usepackage{amssymb}
\usepackage{longtable}
\usepackage{array}
\usepackage{calc}
\usepackage{multirow}
\usepackage{siunitx}
\usepackage{booktabs}
\usepackage{graphicx}
\usepackage{textcomp}
\usepackage[colorlinks=true, linkcolor=blue, citecolor=blue, urlcolor=blue]{hyperref}
\usepackage{setspace}

\begin{document}
	\title{Determining the degree of randomness in multiple-choice question distractors}
\author{J. P. W. Diener}
\email{dienerj@biust.ac.bw}
\affiliation{Department of Physics and Astronomy, Botswana International University of Science and Technology (BIUST), Private Bag 16, Palapye, Botswana}
\author{B.L. Frick}
\affiliation{Centre for Higher and Adult Education, Stellenbosch University, Stellenbosch, South Africa}
\affiliation{Stellenbosch University DSI/NRF Centre of Excellence in STI Policy, Stellenbosch University, Stellenbosch, South Africa}
\author{J. Kriek}
\affiliation{Department of Physics, University of South Africa, South Africa}
\begin{abstract}
Multiple-choice questions are a staple of educational assessment due to their efficiency, but their diagnostic ability is inherently limited. Traditional psychometric techniques infer misconceptions by fitting models to predetermined data patterns, making them ineffective for analysing small, novel, or non-standardised samples where such prior trends are absent. To address this, we introduce a new distractor analysis method based on Information Theory, which treats the randomness in incorrect answers not as noise but as a measurable signal in student answer patterns. By examining the entropy of distractor choices alongside overall accuracy, our approach offers a sample-sized-independent framework for classifying performance. Cross-referencing low-scoring item-level response data against independently documented misconceptions on a standard mechanics inventory shows that, for items identified as most concentrated (lowest degree of randomness), a majority of incorrect responses select the exact response coded to that item's named, interview-validated misconception. Furthermore, items independently flagged elsewhere as diagnostically unreliable are exactly the items the measure identifies as having the highest degree of randomness. The pedagogically actionable cases are those with low degree of randomness in a class's incorrect answers, signalling a shared, re-teachable misconception, whereas a high degree of randomness chiefly serves as the null case against which that concentration is judged. Our analysis is intended as a fast, single-administration triage step: a practical, theory-driven complement to established distractor-analysis methods that helps instructors flag which items warrant closer, more resource-intensive investigation.
\end{abstract}
\maketitle
\begin{figure}[htbp]
	\centering
	\includegraphics[width=0.8\linewidth]{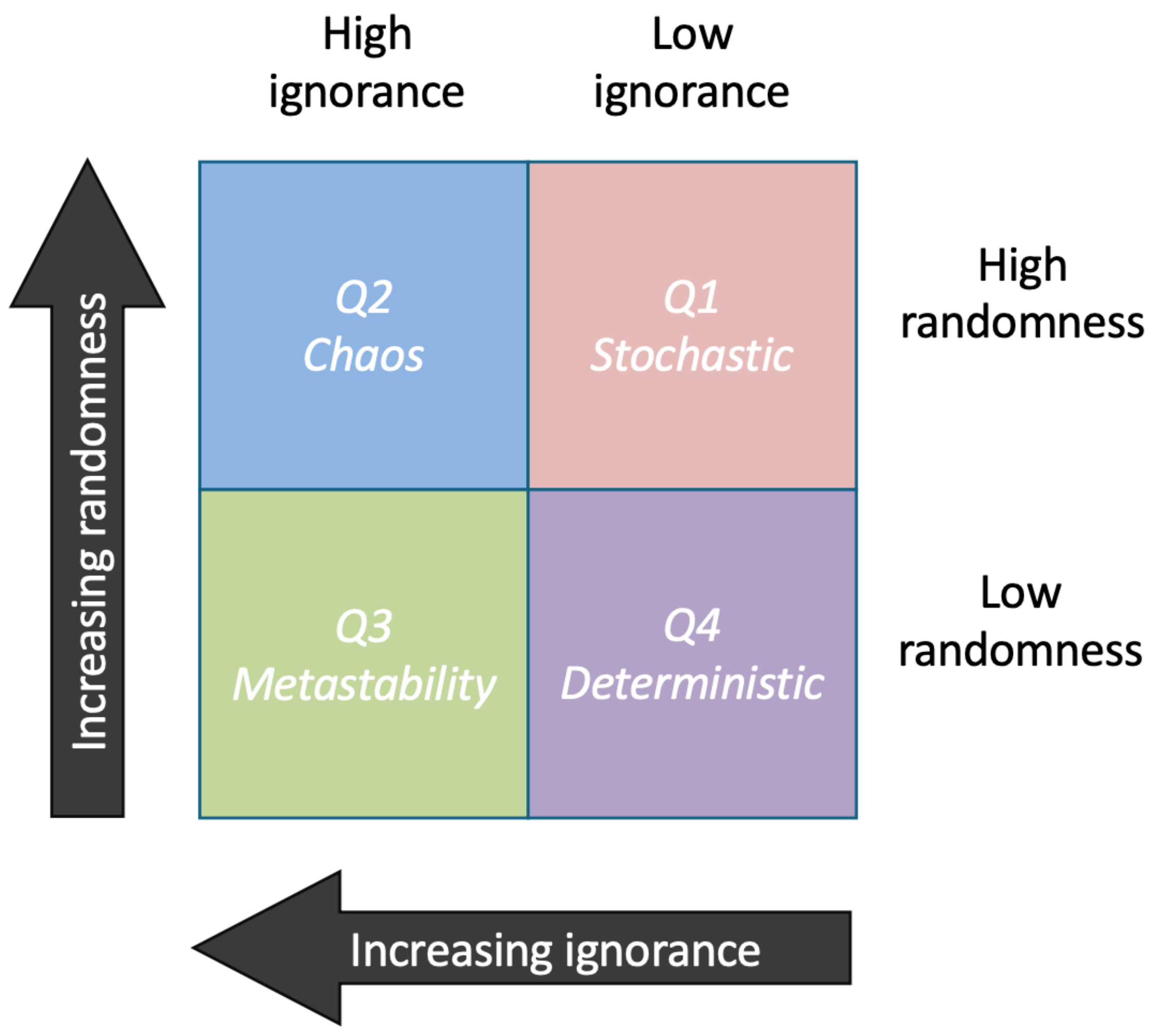}
	\caption{The landscape of predictability}
	\label{fig:1}
\end{figure}
\begin{figure}[htbp]
	\centering
	\includegraphics[width=0.8\linewidth]{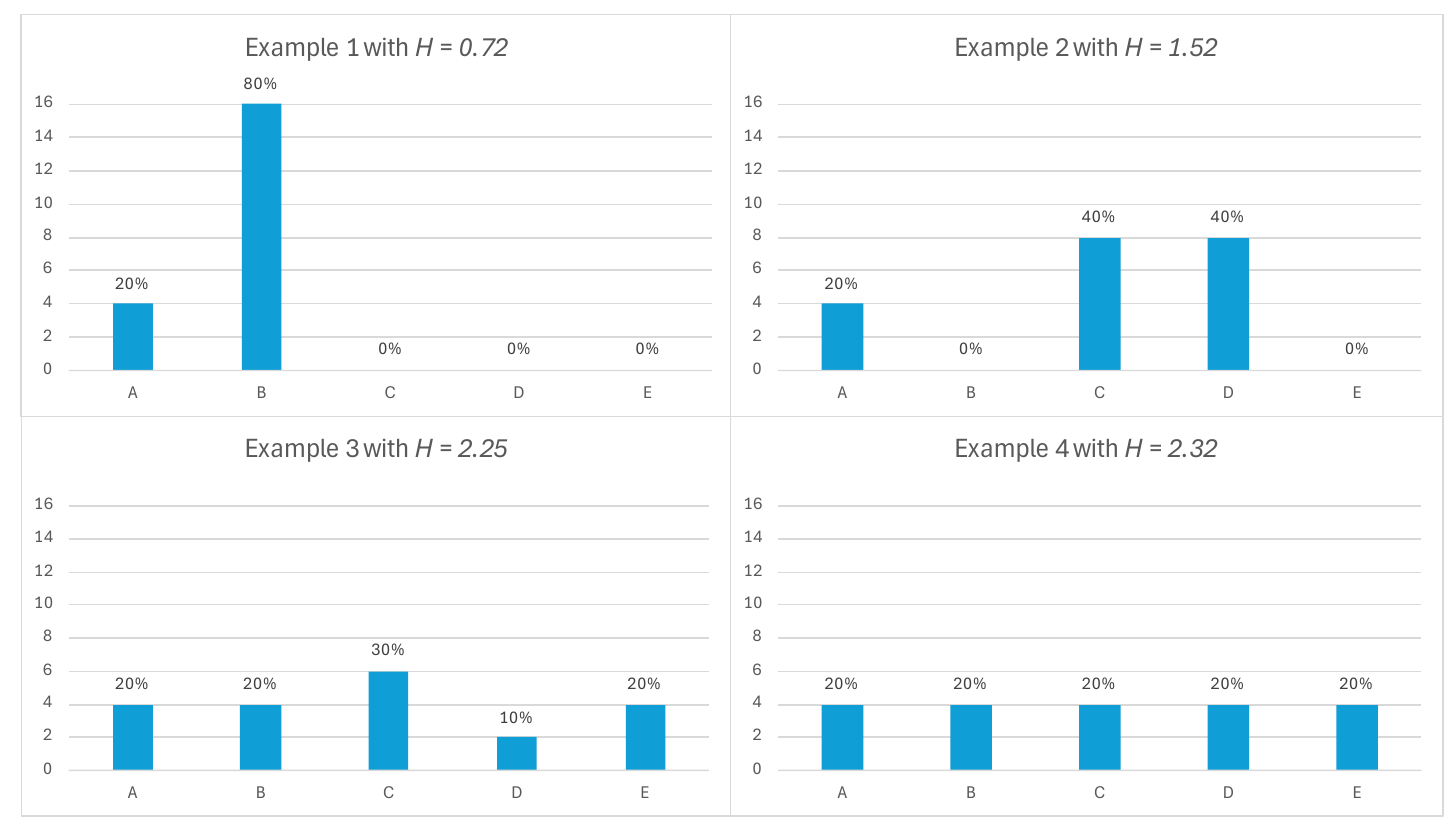}
	\caption{Example MCQ answer frequency distributions along with the Shannon entropy of each example distribution}
	\label{fig:2}
\end{figure}

\hypertarget{introduction}{%
	\section{Introduction}\label{introduction}}
Multiple-choice (MC) items---commonly presented as multiple-choice questions (MCQs), the term used throughout this paper---are a fundamental part of assessment in large-scale educational settings, valued for their efficiency, objectivity, and scalability. However, this efficiency often comes with a drawback: a significant diagnostic challenge. When a class performs poorly on an MCQ assessment, the resulting score or percentage offers a clear measure of failure but provides little understanding of its underlying cause. Instructors are left with a crucial, unanswered question: Are students failing because of shared, coherent misunderstandings that require conceptual re-teaching, or are they merely guessing, which points to a fundamental and widespread lack of engagement or prerequisite knowledge? This dilemma arises from the complex nature of MCQ responses, which the literature shows to be a probabilistic mix of knowledge, strategy, and noise \cite{abu2023,briggs2006,Gierl2017}.\\
\\
It is worth stating at the outset what kind of diagnostic question this paper addresses, since it differs from the question most distractor-analysis methods are built to answer. Methods grounded in item response theory (IRT), classical test theory (CTT), or concentration analysis \cite{bao2001} are primarily oriented toward instrument validation: is this item, or this test, functioning as intended, and should it be revised or retained? That question is answered after the fact, typically with a validated instrument, an established taxonomy of misconceptions, and a sizeable, stable sample to condition on. The question we address is different in kind: given a single classroom assessment, in a single class, what should the instructor do differently in the next lesson? This formative question has to be answerable immediately, from a single administration, often of an assessment nobody has validated. Our analysis is proposed as a fast, low-infrastructure triage step for that situation---one that flags which items show a response pattern coherent enough to reward a targeted re-teach, as a complement to, rather than a replacement for, the heavier validation-oriented methods that remain the right tool once an item has been flagged as worth closer investigation. The analysis draws on principles well-established in decision science, machine learning, and risk analysis—fields that routinely reason under uncertainty—though these perspectives remain underexplored in educational measurement. Unfamiliarity in this context does not imply irrelevance; rather, we adapt these principles to serve the time-pressed classroom instructor.
\hypertarget{epistemic-and-aleatoric-uncertainty}{%
	\subsection{Epistemic and aleatoric uncertainty}\label{epistemic-and-aleatoric-uncertainty}}
An epistemic uncertainty indicates a level of belief or ignorance about a specific but unknown outcome or parameter. This type of uncertainty can be reduced and is therefore diminished by gathering additional information, enhancing models, or developing a deeper understanding \cite{DerKiureghian2009,Fox2011}. In contrast, aleatoric uncertainty arises from the inherent randomness or variability within a phenomenon itself. It is irreducible and persists even with perfect knowledge, as the outcome is fundamentally unpredictable.\\
\\
The distinction between epistemic and aleatoric uncertainty is well established in risk analysis, reliability engineering, and decision theory \cite{DerKiureghian2009,Fox2011}, but it is rarely articulated in these terms in educational measurement. There, the two are seldom held apart: guessing, distraction, and response noise are typically treated as a nuisance to be modelled and subtracted away so that a cleaner estimate of the underlying construct—ability, knowledge, the epistemic quantity—can be recovered. IRT's guessing parameter and the slip and guess parameters of cognitive diagnostic models (CDM) are typical of this stance: guessing is something to be corrected for on the way to a better ability estimate, not a diagnostic quantity worth measuring in its own right \cite{Han2012}. \citet{Han2012} showed that the $c$-parameter of the 3PL IRT model is often estimated with substantial bias and instability, functioning primarily as a statistical correction rather than a substantive indicator of behavior. \citet{Hsu2020} similarly model rapid guessing in CDM  as a distortion to be removed, equating it with guessing with no required attributes. This stance is understandable for validation-oriented models, but it overlooks systematic convergence on distractors—a pattern that may be pedagogically revealing.\\
\\
The combination of aleatoric and epistemic uncertainties can be seen as varying levels of randomness (aleatoric uncertainty) and ignorance (epistemic uncertainty), as shown in Figure~\ref{fig:1}. The four quadrants illustrate the landscape of predictability. The first quadrant (Q1) features high randomness but low ignorance. In this area, the statistical likelihood of outcomes is well known, but individual results still remain unpredictable---a state of measured uncertainty. Conversely, Q2 exhibits both high randomness and high ignorance---a domain of complete unpredictability where neither outcomes nor their probabilities can be determined. Q3 presents a more misleading picture: low randomness but high ignorance. This indicates surface-level stability, but the significant lack of understanding makes the system's true behavior unreliable and opaque. Finally, Q4, where both randomness and ignorance are low, describes a deterministic regime. With minimal noise and high clarity, outcomes become reliably predictable under known conditions. Thus, the interplay of uncertainties is highlighted: low randomness does not necessarily mean predictability (Q3), while low ignorance does not always imply predictability (Q1).
\\
\\
In everyday classroom terms: a high-scoring student who avoids the distractors is behaving predictably (low randomness, low ignorance); a low-scoring student whose wrong answers scatter evenly across every distractor is guessing (high randomness); and a low-scoring student whose wrong answers cluster on one particular distractor is making a consistent, and therefore potentially identifiable, error (low randomness despite low performance). The formal quadrant language above is used only to make this distinction precise enough to calculate.\\
\\
In educational assessment, an examinee's score is influenced not only by their knowledge but also by aleatory uncertainty---the inherent randomness of question sampling and guessing \cite{burton2001}---and epistemic uncertainty---their subjective confidence and risk preferences shaped by the testing context \cite{bereby2002,budescu2015}.\\
\\
The item validation focus of the currently available psychometric models creates a practical gap for the typical instructor. In the classroom---especially in non-standardised, lower-stakes, or smaller-sample contexts---the powerful tools of formal psychometrics and decision theory are often out of reach or not taken into account. Classical distractor analysis may identify non-functioning options, and item difficulty/discrimination indices can highlight problematic questions, but these methods struggle to formally quantify the level of randomness in student errors. They cannot reliably distinguish whether a pattern of incorrect answers indicates systematic epistemic failure (like a shared misconception) or random noise (widespread guessing). Therefore, instructors might waste pedagogical effort analyzing and re-teaching a mistaken misconception when the real intervention needed is much more fundamental.
\hypertarget{problem-statement}{%
	\subsection{Problem statement}\label{problem-statement}}
MCQs are efficient for scoring but poor for classroom diagnosis. Poor results lack actionable insights, as traditional analysis needs large datasets and provides only general, non-specific feedback \cite{Crocker2008,Nunnally1994}.\\
\\
Current approaches suffer from three critical limitations. First, traditional analysis yields only aggregate, non-specific metrics---such as item difficulty and discrimination indices---that indicate whether a question was challenging or differentiated between high and low performers, but do not explain why students selected incorrect answers \cite{delatorre2014}. Second, treating incorrect responses as a uniform category (dichotomous scoring) obscures the essential distinction between systematic misunderstanding (where errors cluster around specific distractors, revealing targetable misconceptions) and unstructured guessing or fragmented knowledge (where errors distribute randomly, indicating a fundamental lack of engagement or coherence) \cite{abu2023,dubins2016}. Third, while more advanced methods exist these approaches typically require substantial sample sizes, strong a priori cognitive theories, and complex calibration \cite{briggs2006,Gierl2017}. Moreover, probabilities are conceptualized as epistemic entities---denoting degrees of belief or propensities grounded in modelled relationships---rather than as direct estimators of aleatoric uncertainty, that is, the intrinsic randomness inherent in response patterns \cite{Downing2006,haladyna2013}.\\
\\
Consequently, instructors face a persistent diagnostic dilemma. They may invest considerable pedagogical effort in deconstructing and re-teaching what appears to be a shared misconception, when in reality the poor performance stems from widespread guessing or deeply fragmented knowledge that requires a fundamentally different intervention---such as a foundational review or activities to build conceptual coherence. This inefficiency stems from a key unmet need: the lack of a practical, classroom-ready tool that can reliably measure the aleatoric uncertainty in student error patterns to distinguish between structured misunderstanding and unstructured noise.\\
\\
This paper, therefore, addresses the following core problem: While existing psychometric methods excel at validating tests and diagnosing predefined knowledge gaps, there is a lack of simple, robust, and sample-size-independent mechanisms to diagnose the nature of collective failure---specifically, to differentiate between systematic class-wide misconceptions, knowledge fragmentation, or unstructured guessing. We propose that the solution lies in applying an aleatoric lens grounded in Information Theory. This leads to our central research question: \emph{How does an information-theoretic measure of uncertainty in distractor choice deliver fast, practical, and robust diagnostic feedback for MCQ-based instruction---overcoming a core weakness of conventional psychometrics in classroom settings?}\\
\\
The rest of the paper is organised as follows: it starts with an introduction to the proposed analysis method and demonstrates its application to sample data to illustrate how it can be used. Next, the method is applied to collected data in a low-stakes conceptual assessment, and the implications are discussed. In the final section, potential limitations are identified and examined.
\hypertarget{information-theoretical-analysis}{%
	\section{Information Theoretical Analysis}\label{information-theoretical-analysis}}
The method presented here is called distractor Information Theoretical Analysis (dITA). dITA relies on similar quantities as those found in Information Theory, the mathematical theory of communication \cite{shannon1948}.
\hypertarget{information-theory}{%
	\subsection{Information Theory}\label{information-theory}}
All communication involves exchanging a message that contains specific information. Information Theory does not focus on whether the transmitted information is correct or incorrect, but rather on whether any information is lost during transmission. The main assumption is that any message (information being transmitted) is chosen from the set of all possible messages that could be transmitted \cite{shannon1948}. Therefore, a message can be classified based on the probability of its selection from the complete set of possible messages. More precisely, the classification of a message relies on its \emph{entropy}. Entropy is a concept originally linked to Physics, especially statistical physics, and can be regarded as a measure of unexpectedness or the degree of order/disorder (randomness) within a system.~\\
\\
In Information Theory, the Shannon entropy quantifies the measure of uncertainty associated with a random variable \cite{cover2006}. The Shannon entropy, $H$, is
\begin{equation}
	H = - \sum_{i}^{}p_{i}\log_{2}\left( p_{i} \right),
	\label{eq:1}
\end{equation}
where
\begin{itemize}
	\item
	$p_{i}$ is the probability of information associated with the $i$\textsuperscript{th} information state, and,
	\item
	$i$ labels the information states.
\end{itemize}
An example of the $i$\textsuperscript{th} information state could be the $i$\textsuperscript{th} letter of a word in a sentence. The letter can be chosen from the alphabet, but in the context of the sentence it could also possibly be a space or punctuation mark. The probability $p_{i}$ refers to a specific letter that occupies the $i$\textsuperscript{th} position being selected from the set of all possible options. For brevity, the Shannon entropy represented by $H$, will be referred to as the entropy.
\hypertarget{rationale-of-applying-information-theory-to-mcq-analysis}{%
	\subsection{Rationale of applying information theory to MCQ analysis}\label{rationale-of-applying-information-theory-to-mcq-analysis}}
The epistemic/aleatoric split introduced above has a direct consequence for which statistical object is the right one to compute. IRT and CTT are built to estimate the epistemic quantity: a point estimate of a student's or a class's underlying ability, refined by treating guessing as error to be averaged out. The aleatoric quantity, by contrast, is not a property of any single point estimate but a property of the \emph{shape} of the response distribution itself, at whatever level of ability the class happens to have---how many ways, and how evenly, a class distributes its wrong answers. Shannon entropy is the natural formal object for exactly this quantity: it is defined on a probability distribution rather than on a point estimate and measures how surprising a specific outcome is; it is maximised precisely when that distribution is uniform (complete unpredictability among the distractors), and it is minimised when the distribution collapses onto a single option (complete, albeit incorrect, consistency). Where IRT and CTT ask how much a student or class knows, entropy asks a differently shaped question: given that they did not know, how uniformly did that not-knowing express itself? The second question is the aleatoric one, and it is the one the remainder of this paper operationalises.\\
\\
When applied to MCQ assessments, entropy measures the level of uncertainty within a system, which in this case is the answer to the MCQ. When applied to incorrect answers, the distractor choices, it can expose how diverse the choice in distractors were. High entropy in incorrect answers may suggest that participants choose a variety of distractor options, indicating they make different kinds of errors. Conversely, low entropy might imply that participants tend to make the same type of mistake, reflected by a more uniform selection of distractors.\\
\\
One way to analyze these MCQ assessment results is to plot the frequency distributions of how many times each answer was selected. Four examples of such plots are shown in Figure~\ref{fig:2}.
All plots show twenty participants responding to a hypothetical MCQ with four distractors and one correct answer. Assuming ``A'' is the correct option for all examples, only $20$\% of the participants chose the correct answer, while the rest distributed their answers among the distractors. Each example in Figure~\ref{fig:2} illustrates a different hypothetical distractor answer distribution. Comparing examples 1 and 4 illustrates the extreme cases: In example 1, all incorrect participants chose the distractor ``B'', whereas in example 4, the participants spread their incorrect answers evenly among all options. In example 1, all incorrect participants chose the same distractor. Hence, there is probably a more easily identifiable misconception to be addressed, whereas, for example, scenario 4, it would be more complicated.\\
\\
While the distributions in Figure~\ref{fig:2} can be visually distinguished using the frequency plots, applying this method to the entire assessment involves creating a separate plot for each question. Instead of plotting such a frequency distribution, its Shannon entropy, $H$, can be calculated. $H$ measures the spread within the distribution, enabling simple comparison between frequency plots. However, because $H$ is a dimensionless measure, it does not inherently provide a meaningful scale for comparing distributions. This limitation prompts a follow-up question: How can the randomness in students\textquotesingle{} selection of incorrect answers in MCQs be quantified for comparison?
\hypertarget{relation-to-concentration-based-distractor-analysis}{%
	\subsection{Relation to concentration-based distractor analysis}\label{relation-to-concentration-based-distractor-analysis}}
The closest existing precedent to the approach proposed here is concentration analysis, developed in physics education research and applied to the Force Concept Inventory (FCI): \citet{bao2001} define a concentration factor, $C$, and a score-independent concentration deviation, $\Gamma$, both built from the sum of squared response proportions across options---a Simpson- or Herfindahl-type dispersion index \footnote{Sum of squares indices used in ecology and economics respectively.}---and combine $\Gamma$ with item score on a two-dimensional plot to classify items as reflecting a single dominant misconception, competing misconceptions, or near-random guessing. This is structurally the same diagnostic move as we applied: a score-independent dispersion statistic over distractor choices, read jointly with performance. The two methods differ in the family of dispersion index used---sum-of-squares versus Shannon entropy---which is not a neutral choice: a sum-of-squares statistic is most sensitive to the size of the single largest wrong-answer cluster, whereas Shannon entropy is more sensitive to how many distinct wrong answers are in play at all, so the two measures can rank the same pair of response distributions in opposite order when their underlying distributions are not simple rescalings of one another. For our purpose of supplying actionable pedagogical information entropy's broader sensitivity to the number of distinct error types is deemed to be the better approach.\\
\\
Concentration analysis and dITA also differ in the infrastructure they presuppose. Bao and Redish's interpretation of a concentrated response as a specific misconception rests on qualitative interview research that had already catalogued those misconceptions \cite{bao2001}. Our analysis is intended for the more common classroom situation in which no such qualitative catalogue or multi-class sample exists---an instructor's own, possibly one-off, classroom quiz. We return to this distinction, and to full-information item analysis \cite{haladyna2021} as a further point of comparison, in the Discussion.
\hypertarget{method}{%
	\section{Method}\label{method}}
As stated in Section \ref{information-theoretical-analysis}, the method presented here is called distractor Information Theoretical Analysis (dITA). In dITA the degree of randomness of incorrect answers to an MCQ is quantified based on the simple equiprobability definition of randomness \cite{batanero2015} where all answer options are equally likely. The analysis starts by calculating the entropy of the incorrect (distractor) responses in a hypothetical scenario of such equiprobable randomness, where the incorrect options are selected. The entropy corresponding to this scenario is denoted by $dH$. This value is then compared to the entropy of the actual collected data, distinguished by adding a prime: $dH'$. By comparing the entropy of equiprobable, and thus random, distractor selection with that of the observed data, a single parameter can be derived to quantify the degree of randomness present in the collected responses.\\
\\
The first step in the motivation of this methodology is to consider the dichotomous distinction between correct and incorrect answers: For a class of $N$ students, let's assume $N_{K}$ of those $N$ chose the key as their (correct) answers to the MCQ. That leaves $N_{D} = N - N_{K}$ students who gave an incorrect answer.\\
\\
To describe the behavior of the class of $N$ students who submitted an answer to the MCQ, the probability that any student in the class submitted the correct answer is $p_{K} = \frac{N_{K}}{N}$, and correspondingly, the probability that a student picked at random submitted an incorrect answer is $p_{D} = \frac{N_{D}}{N}$. Furthermore, since $N_{D} + N_{K} = N$, it also means that $p_{D} = 1 - p_{K}$.
\hypertarget{equiprobable-choice-of-distractor}{%
	\subsection{Equiprobable choice of distractor}\label{equiprobable-choice-of-distractor}}
For the $N_{D}$ students who chose an incorrect answer, if their choice of distractor was made at random, it would imply that the probability of choosing a specific distractor is equal to the probability of choosing any other distractor. Therefore, if the number of distractor options is $D$, the probability of having a randomly chosen incorrect answer is, for $N$ students where the probability of giving the correct answer is $p_{K}$, the probability of choosing the incorrect answer at random out of $D$ distractor options is $\frac{p_{D}}{D} = \frac{(1 - p_{K})}{D}$.\\
\\
All MCQ answer options thus have a probability associated with them, and the Shannon entropy associated with each state can be calculated. Since our concern is the incorrect answers, the entropy associated with the incorrect answers are equally spread between all $D$ distractor options (which we will called $dH$) is
\begin{equation}
	dH = - \sum_{i = 1}^{D}p_{i}\log_{2}\left( p_{i} \right) = - \left( 1 - p_{K} \right)\log_{2}\left( \frac{1 - p_{K}}{D} \right).
	\label{eq:2}
\end{equation}

It is worthwhile to note that the above expression only depends on the probability $p_{K}$ and the number of distractor options $D$ of the MCQ. Therefore, if the percentage of correct answers to an MCQ is known, in addition to the number of distractors, then $dH$ can be calculated. However, knowing the value of $dH$ is not very insightful since, as with most entropy calculations, the quantity of interest is not the absolute entropy but rather the relative entropies of different system configurations.\\
\\
Throughout, $D$ denotes the number of available distractor options, not only the functioning distractor options---those selected by at least one student. Validity of distractors are not a concern, emphasising the complementary nature of our approach as well as the importance of the established IRT and CTT analysis to determining functioning distractors. \\
\\
Our aim is thus to compare the $dH$ calculated from Equation~\eqref{eq:2}, where it is assumed that when an incorrect answer was submitted choice of distractor options was made at random, to the entropy of the collected incorrect answers data, $dH'$.
%\\
%
\hypertarget{entropy-of-incorrect-answers}{%
	\subsection{Entropy of incorrect answers}\label{entropy-of-incorrect-answers}}
The entropy of the collected incorrect answer data $dH'$ is calculated based on Equation~\eqref{eq:2}, but with some slight modifications:
\begin{equation}
	dH' = - \sum_{i = 1}^{D}{p_{i}'}\log_{2}\left( p_{i}' \right).\
	\label{eq:3}
\end{equation}
 Equation~\eqref{eq:3} is similar to the first part of Equation~\eqref{eq:2}, but in Equation~\eqref{eq:3} the probabilities associated with each of the $D$ distractor options, $p_{i}'$, are based on the collected data of the submitted answers and not the assumption of an equiprobable choice of distractor as in Equation~\eqref{eq:2}.

\hypertarget{analysis-assumption}{%
	\subsection{Analysis assumption}\label{analysis-assumption}}
The fundamental assumption on which this analysis is based is that the calculated $dH^{'}$, from Equation~\eqref{eq:3}, is compared with $dH$ from Equation~\eqref{eq:2}. In this comparison $dH'$ from  Equation~\eqref{eq:3} is set equal to the modified version of Equation~\eqref{eq:2}, where $D$ has been replaced with $D'$:
\begin{equation}
	{dH}^{'} = - \sum_{i = 1}^{D}{p_{i}'}\log_{2}\left( p_{i}' \right) = - \left( 1 - p_{K} \right)\log_{2}\left( \frac{1 - p_{K}}{D^{'}} \right),
	\label{eq:4}
\end{equation}
This substitution allows for a comparison between $dH'$ and $dH$. However, rather than a direct comparison of the entropies $dH'$ and $dH$, the values of $D'$ is compared to $D$.

\hypertarget{estimating-randomness-in-distractor-answers}{%
	\subsection{Estimating randomness in distractor answers}\label{estimating-randomness-in-distractor-answers}}
Combining the descriptions and motivation of Equations~\eqref{eq:2} and \eqref{eq:3}, $D'$ could be interpreted as the \emph{mimicked number of random (equiprobable) distractors} that corresponds to $dH'$. If $dH' = dH$, it follows automatically that $D' = D.$ However, in the instances where students who submitted incorrect answers did not make the choice of distractor randomly, then $D' \neq D$.\\
\\
The defining equation for $D'$ is obtained from Equation~\eqref{eq:4} as:
\begin{equation}
	D^{'} = \left( 1 - p_{K} \right)2^{\frac{dH'}{\left( 1 - p_{K} \right)}}.
	\label{eq:5}
\end{equation}
Equation (\ref{eq:5}) may look complicated, but it can be easily calculated using a spreadsheet-based computer programme or mathematical package \cite{DienerGitHub}.\\
\\
An equivalent, and computationally simpler, closed form for $D'$ follows directly from Equations~(\ref{eq:2})--(\ref{eq:5}). Writing $p_i = n_i/N_D$ for the distribution of incorrect responses \emph{conditional} on being incorrect (where $N_D = N(1-p_K)$ is the number of incorrect respondents), the $p_K$-dependent terms cancel exactly, leaving
\begin{equation}
	D' = 2^{H_1(p)}, \qquad H_1(p) = -\sum_{i=1}^{D} p_i \log_2 p_i.
	\label{eq:5b}
\end{equation}
$D'$ is therefore the effective number of distractors in use---the Hill number of order 1 \cite{Hill1973} of the incorrect-response distribution alone---with no remaining dependence on $p_K$.  This closed form, and its relationship to $\Gamma$ as the corresponding Hill number of order 2, is developed further in an upcoming companion paper.
\subsection{\texorpdfstring{Scaling of $D^{'}$}{Scaling of D\^{}\{\textquotesingle\}}\label{scaling-of-d}}
\begin{figure}[htbp]
	\centering
	\includegraphics[width=0.8\linewidth]{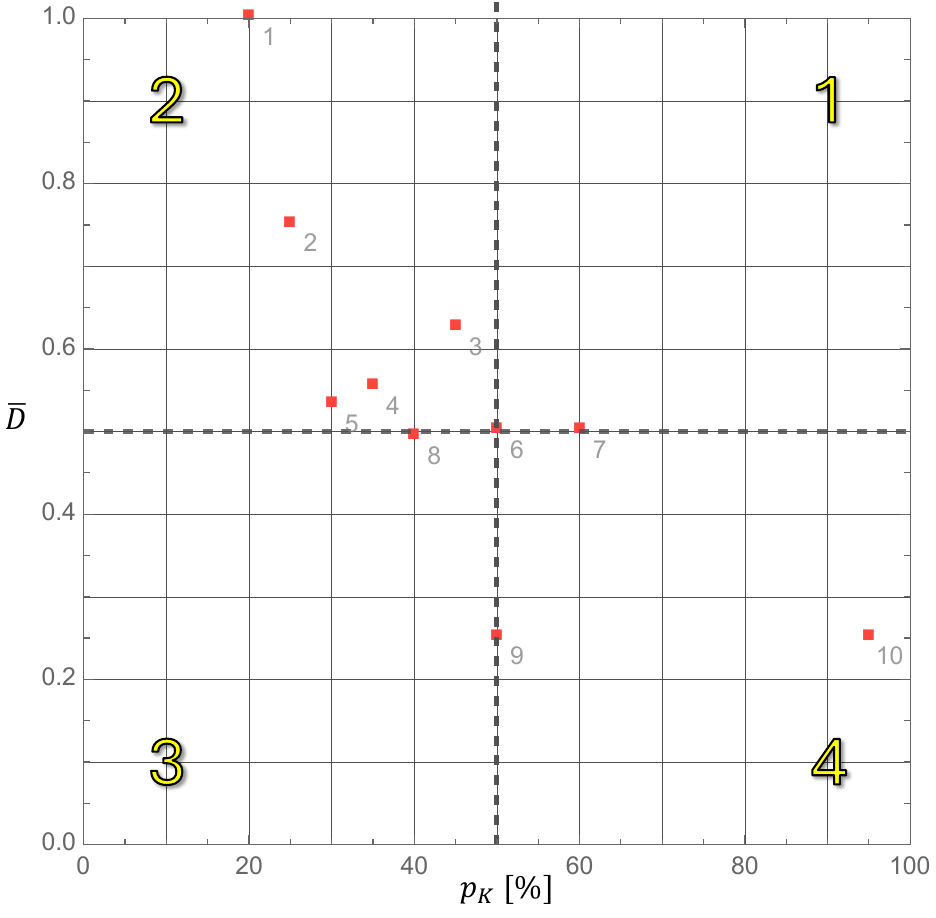}
%	\fbox{\parbox[c][5cm][c]{0.75\linewidth}{\centering\itshape [Figure 3 placeholder -- original: image7]}}
	\caption{The $p_{K}[\%]$-$\bar{D}$ of Table \ref{tab:1}, indicating the row numbers.}
	\label{fig:3}
\end{figure}

Calculating $D'$ from the (submitted) data provides a way to quantify the number of incorrect answer states that students submitted. However, unless all questions in the assessment have the same number of distractor options, $D'$ does not lend itself to comparing different questions in the assessment.\\
\\
This can be addressed by defining a relative quantity that scales a MCQ's $D'$ relative to the actual number of distractor options $D$. This quantity is called $\bar{D}$ (``scaled $D$'') and is
\begin{equation}
	\bar{D} = \frac{D^{'}}{D}.
	\label{eq:6}
\end{equation}
At first glance, the range of $\bar{D}$ would be a fraction between$\ 0$ and $1$ (since $D'$ cannot be greater than $D$). However, the practical range of $D^{'}$ are from $1$ to $D$ and correspondingly $\bar{D}$ would have the practical limits of
\begin{equation}
	\bar{D} \rightarrow 1\text{ as }D^{'} \rightarrow D\text{, and,}\ \bar{D} \rightarrow \frac{1}{D}\ \text{as}\ D^{'} \rightarrow 1.
	\label{eq:7}
\end{equation}
\hypertarget{values-of-mathbfd}{%
	\subsection{\texorpdfstring{Comparison of $\mathbf{\bar{D}}$ to $\mathbf{\Gamma}$}{Comparison of \textbackslash mathbf\{$\bar{D}$\} to \textbackslash mathbf\{$\Gamma$\}}}\label{values-of-mathbfd}}
Table~\ref{tab:1} lists the values of $D, D',$ and $\bar{D}$ for a range of example answer combinations, to illustrate the range and variation in these values. For these distributions, there are multiple instances of similar values of $D^{'}$ and $\bar{D}$ that correspond to completely different answer distributions; at first glance, the results appear to be highly ambiguous. However, if the $\bar{D}$ data is considered in conjunction with the corresponding $p_{K}$ values, clear distinctions emerge (see Figure~\ref{fig:3}). As a comparison to other approaches, the values $\Gamma$ \cite{bao2001} for the distributions are included.\\
\\
\begin{table}[htbp]
	\centering
	\caption{MCQ Answer Options and Statistics}\label{tab:1}
	\begin{tabular}{|c|c|c|c|c|c|c|c|c|c|c|c|c|}
		\hline
%		\multicolumn{1}{|c|}{Descriptors} &\multicolumn{1}{c|}{\#} 
%		& \multicolumn{5}{c|}{MCQ answer options} & \multicolumn{1}{c|}{Total} & $p_K$ & $D$ & $D'$ & $\bar{D}$ & $\Gamma$\\
		\multirow{2}*{Descriptors} & \multirow{2}*{\textbf{\#}} & \multicolumn{5}{c|}{MCQ answer options}&\multirow{2}*{Total} & \multirow{2}*{$p_K$} & \multirow{2}*{$D$} & \multirow{2}*{$D'$} & \multirow{2}*{$\bar{D}$} & \multirow{2}*{$\Gamma$} \\
		\cline{3-7}
		& & $\ $ A $\ $&$\ $ B $\ $&$\ $ C $\ $ & $\ $ D $\ $& $\ $ E $\ $& & & & & & \\
		\hline
%				\# & Key & Distractor 1 & Distractor 2 & Distractor 3 & Distractor 4 & & & & & \\
		\multirow{7}{*}{\rotatebox[origin=c]{90}{Number of submissions\ \ \ } }
	&\textbf{1} & 4  & 4  & 4  & 4  & 4  & 20 & 0.20 & 4 & 4.00 & 1.00 & 0.00 \\
	&	\textbf{2}  & 5  & 5  & 5  & 5  & 0  & 20 & 0.25 & 4 & 3.00 & 0.75 & 0.15 \\
	&\textbf{3}  & 9  & 1  & 6  & 4  & 0  & 20 & 0.45 & 4 & 2.50 & 0.63 & 0.32 \\
	&	\textbf{4}  & 7  & 10 & 1  & 1  & 1  & 20 & 0.35 & 4 & 2.21 & 0.55 & 0.56 \\
	&	\textbf{5}  & 6  & 11 & 1  & 1  & 1  & 20 & 0.30 & 4 & 2.13 & 0.53 & 0.59 \\
	&	\textbf{6} & 10 & 5  & 5  & 0  & 0  & 20 & 0.50 & 4 & 2.00 & 0.50 & 0.41 \\
	&	\textbf{7}  & 12 & 0  & 4  & 4  & 0  & 20 & 0.60 & 4 & 2.00 & 0.50 & 0.41 \\
	&	\textbf{8}& 8  & 0  & 7  & 5  & 0  & 20 & 0.40 & 4 & 1.97 & 0.49 & 0.43 \\
	&	\textbf{9}  & 10 & 0  & 10 & 0  & 0  & 20 & 0.50 & 4 & 1.00 & 0.25 & 1.00 \\
	&	\textbf{10} & 19 & 0  & 1  & 0  & 0  & 20 & 0.95 & 4 & 1.00 & 0.25 & 1.00 \\
		\hline
	\end{tabular}
\end{table}
Comparison of rows 5 and 8 allows a direct comparison between $\Gamma$ and $\bar{D}$. Both rows have the same four nominal distractor options available (14 incorrect submissions for Row 5, 12 for Row 8), but where Row 8 concentrates its incorrect submissions on just two of the four options (C and D), Row 5 places a single, larger plurality on one option while still touching all three remaining options. $\Gamma$, being most sensitive to the size of the single largest distractor share, ranks Row 5 ($\Gamma=0.59$) as more concentrated than Row 8 ($\Gamma=0.43$). $\bar{D}$, being more sensitive to how many distinct options are actually in use, ranks them the other way around: Row 5 ($\bar{D}=0.53$) as \emph{more} random than Row 8 ($\bar{D}=0.49$), since Row 5's incorrect submissions are distributed over all four options while Row 8's are confined to two. This is the same divergence discussed conceptually in Section~\ref{relation-to-concentration-based-distractor-analysis}, reproduced here directly within the illustrative data already introduced in this section, rather than as an abstract case.\\
\\
Rows 4 and 6 illustrate the same divergence in the other direction: a case where $\Gamma$ is arguably the more appropriate statistic to consult. In Row 4, one option captures 10 of the 13 incorrect submissions  (77\%), with the remaining three submissions scattered one apiece across the other options; in Row 6, incorrect submissions split evenly, 5 and 5, between exactly two options, with the remaining two untouched. $\Gamma$ correctly distinguishes these as different in kind: Row 4 ($\Gamma=0.56$) is read as substantially more concentrated than Row 6 ($\Gamma=0.41$), consistent with Row 4 having a single, clearly dominant incorrect answer and Row 6 having two competing ones of equal weight---the same one-model versus two-model distinction \citet{bao2001} build their own concentration measure around. $\bar{D}$ ranks these the other way ($\bar{D}=0.55$ for Row 4 against $0.50$ for Row 6), penalising Row 4 for the three students who each drifted to a different minor option, even though 77\% of Row 4's incorrect submissions share a single answer. For a question this paper does not centre on---whether an item reliably triggers one identifiable, nameable incorrect answer, as opposed to leaving open which of several competing answers a re-teach would need to address---$\Gamma$'s reading of Rows 4 and 6 is the more useful one, and this is precisely the instrument-validity-oriented sensitivity attributed to $\Gamma$-type measures in Section~\ref{positioning-ita-relative-to-instrument-validation-methods}. Why $\Gamma$ is the more useful statistic specifically for this pair, and not for Rows 5 and 8, is revisited graphically in Section~\ref{graphGD}, once the $p_K$--$\bar{D}$ quadrant scheme has been introduced.\\
\\
The remaining rows populate the range between these two extremes, with instances of similar or the same values of $\bar{D}$. To disentangle their meaning, Figure~\ref{fig:3} shows $\bar{D}$ as a function of the corresponding $p_{K}${[}\%{]} values.

\hypertarget{relevance-of-calculated-quantities}{%
	\section{Relevance of calculated quantities}\label{relevance-of-calculated-quantities}}
To illustrate the relevance of $\bar{D}$ a closer look will be taken at the data in Table~\ref{tab:1} as well as the representation thereof in Figure~\ref{fig:3} where, same as Figure~\ref{fig:1}, the plot area is divided into quadrants. Large values of $p_{K}$ indicate a high percentage of correct answers and thus limited ignorance while small of $p_{K}$ would be indicative of high ignorance. Furthermore, $\bar{D}$ quantifies the degree of randomness thus allowing for a direct comparison to Figure~\ref{fig:1}.

\hypertarget{graphical-interpretation}{%
	\subsection{Graphical interpretation}\label{graphical-interpretation}}
In Figure~\ref{fig:3} Q1 is defined by large values of both $p_{K}$ and $\bar{D}$. Large $\bar{D}$ indicates a tendency towards randomness in the incorrect answers; $\bar{D} \geq 0.5$ implies that more than half of the question's distractors were viable answers; large $p_{K}$ indicates that more students submitted the correct answer than the incorrect answer. Therefore, Q1 represents the situation in which most students submitted the correct answer, while those who did not were more likely to randomly select an incorrect answer, indicative of a stochastic process at work in terms of distractor selection: all possible outcomes are known, but their prevalence is unpredictable.\\
\\
A caveat applies throughout this classification: a distribution close to uniform across the distractors is consistent with random guessing, but it is not proof of it. The same pattern can arise from several equally well-constructed, equally attractive distractors each capturing a different partial-knowledge state, rather than from guessing as such (see also Possible shortcomings). Distinguishing these two explanations from $\bar{D}$ alone is not possible; doing so needs either independent evidence (e.g., interviews, documented misconceptions) or conditioning the response pattern on students' total-score group, which we treat as a direction for future work rather than a claim made by the present analysis.\\
\\
Q2 is defined by low $p_{K}$ and high $\bar{D}$. As with Q1, data points in Q2 indicate a lack of consistency in the choices of incorrect answer options (indicated by high $\bar{D}$ values) and thus a tendency to select answers in what appears to be random selection. However, in contrast to Q1, most of the students did not submit the correct answer (indicated by low $p_{K}$ values). The chaotic combination of high ignorance regarding the correct answer (low values of $p_{K}$) and the tendency to select (incorrect) answers at random, could be indicative of widespread confusion and guessing for questions appearing in Q2.\\
\\
In Q3, most students also submitted the incorrect answer. However, the low $\bar{D}$ values indicate some consistency (lack of randomness) in the incorrect answers. Due to the class's consistency in the choice of distractor/s, there could be an opportunity to identify the reason for the ignorant, but consistent choice of incorrect answers by most of the class.\\
\\
In Q4, most students provided the correct answer, while the low $\bar{D}$ values indicate that those who submitted incorrect answers were aligned with the rest of the class in their choice that submitted an incorrect answer, indicative of more predictable behavior in Q4.\\
\\
Of these four regions, Q2 and Q4--where $p_{K}$ is already high or low, and $\bar{D}$ offers comparatively little additional information--are of secondary interest. The pedagogically consequential contrast is between Q1 and Q3: a high degree of randomness (Q1) is the null case, consistent with undifferentiated guessing among students who did not know the answer, and offers little further purchase for instruction. A low degree of randomness on a poorly answered question (Q3), by contrast, is the diagnostically rich case as it signals that the class converged on a shared, identifiable error, and is therefore the pattern most worth an instructor's attention and most amenable to a targeted pedagogical response. The remainder of this paper accordingly emphasises questions with a low degree of randomness, since these are where the method offers the clearest actionable insight.
\hypertarget{graphGD}{%
	\subsection{\texorpdfstring{Revisiting $\mathbf{\bar{D}}$ and $\mathbf{\Gamma}$ graphically}{Revisiting \textbackslash mathbf\{$\bar{D}$\} and \textbackslash mathbf\{$\Gamma$\} graphically}}\label{graphGD}}
Revisiting the $\bar{D}$--$\Gamma$ divergence from Table~\ref{tab:1} against this quadrant scheme sharpens, rather than merely restates, the case made there for each statistic. Row 5 ($p_K=0.30$, $\bar{D}=0.53$) and Row 4 ($p_K=0.35$, $\bar{D}=0.55$) both fall in Q2; Row 8 ($p_K=0.40$, $\bar{D}=0.49$) falls in Q3; Row 6 ($p_K=0.50$, $\bar{D}=0.50$) sits exactly at the $p_K=\bar{D}=0.5$ boundary shared by all four quadrant definitions.\\
\\
For Rows 5 and 8, the quadrant assignment agrees with, and independently corroborates, $\bar{D}$'s reading. Row 5's position in Q2 places it in this paper's own ``widespread confusion and guessing'' region; Row 8's position in Q3 places it in the diagnostically rich region this paper treats as worth an instructor's attention. Two different tools built into this paper---the raw $\bar{D}$ comparison and the qualitative quadrant classification---reach the same conclusion for this pair without one being derived from the other.\\
\\
For Rows 4 and 6, the quadrant scheme offers no such corroboration, because it has nothing to offer at all: Row 6 sits precisely at the single point where all four quadrants meet. The graphical apparatus this paper relies on throughout is silent exactly where it is asked to distinguish Row 4 from Row 6. This gives a more principled reason for consulting $\Gamma$ here than simply noting that it happens to rank the two rows differently: $\Gamma$ is doing useful work precisely at the point where $\bar{D}$'s own interpretive scheme runs out of resolving power, not as an arbitrary alternative to it. This sharpens, rather than competes with, the instrument-validity framing already given to $\Gamma$-type measures in Section~\ref{positioning-ita-relative-to-instrument-validation-methods}.\\
\\
One caveat is worth stating plainly. Row 6 landing exactly on the four-way boundary is a property of this specific illustrative data, not a general claim that two-competing-model items always sit at that intersection. The broader point---that $\Gamma$'s value is clearest exactly where $\bar{D}$ and its graphical interpretation are least informative---does not depend on that coincidence, but the illustration of it here does.
\hypertarget{identification-of-targeted-questions}{%
	\subsection{Identification of targeted questions}\label{identification-of-targeted-questions}}
Based on an assessment's $p_{K}$-$\bar{D}$ data, diagnostic inferences can be drawn regarding the state of the class's knowledge and/or possible misconceptions or confusion that collectively manifested during the assessment. This could yield actionable data for classroom interventions. An illustration based on low-stakes assessment with low-scoring data is provided next.
\hypertarget{application-to-collected-data}{%
	\section{Application to collected data}\label{application-to-collected-data}}

\begin{figure}[htbp]
	\centering
	\includegraphics[width=0.8\linewidth]{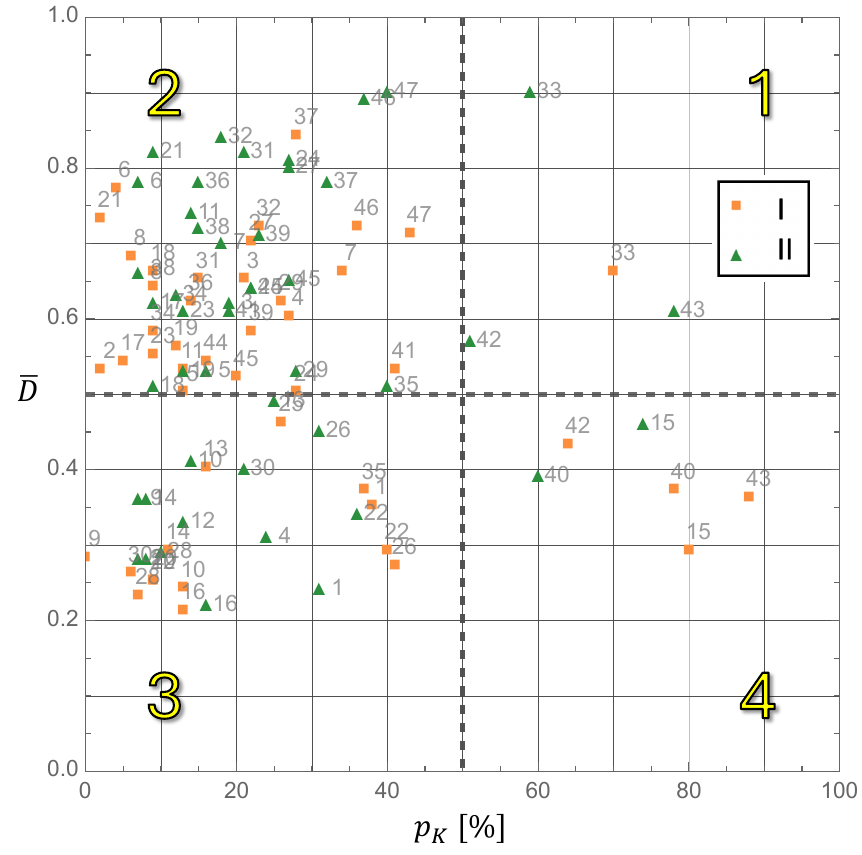}
%	\fbox{\parbox[c][5cm][c]{0.75\linewidth}{\centering\itshape [Figure 4 placeholder -- original: image8]}}
	\caption{The $p_{K}[\%]$-$\bar{D}$ values for the two cohorts.}
	\label{fig:4}
\end{figure}

The experimental data were collected over two academic years (2021/2022 and 2022/2023) from students in an introductory physics course at a university in Botswana using the Force and Motion Conceptual Evaluation (FMCE) \cite{Thornton1998}. Participation was both voluntary as well as anonymous and did not contribute towards the students' mark for the course. No demographic characteristics were collected. Ethical approval and the appropriate research permit to conduct the study were obtained. The data that support the findings of this study are available on request from the corresponding author, but data are not publicly available due to the stipulations of the ethical approval. The participation details are listed in Table~\ref{tab:2}.
\begin{table}[htbp]
	\centering
	\small
	\caption{Summary of PHYS101 test deployments and reliability statistics}
	\label{tab:2}
	\begin{tabular}{lcc}
		\toprule
		& \textbf{2021/2022 (I)}& \textbf{2022/2023 (II)} \\
		\midrule
		Enrolment & 403 & 454 \\
		Date & Dec 2021 & Nov 2022 \\
		Mode &  Online &  Online + Physical \\
		Submissions ($N$)  & 48  & 206 \\
		Cronbach's $\alpha$ & 0.958  & 0.951 \\
		\bottomrule
	\end{tabular}
\end{table}
\hypertarget{data-instrument}{%
	\subsection{Data instrument}\label{data-instrument}}
The FMCE is a physics multiple-choice assessment aimed at high school and introductory tertiary students, focusing on their understanding of the concept of force and its relation to motion, as the name suggests. It purpose is to assess students' understanding of Newtonian mechanics by distinguishing Newtonian and non-Newtonian thinking in students \cite{Thornton1998, madsen2018}. Since its inception, the FMCE has been widely adopted and translated into various languages \cite{pportFCME2025}. Each FMCE question offers between five and nine answer options. Every question has only one correct answer, resulting in four to eight distractors. The use of the FMCE in relation to this work is incidental, as the content of the assessment is not directly relevant to the study, apart from the fact that students should be familiar with it.
\hypertarget{data-manipulations}{%
	\subsection{Data manipulations}\label{data-manipulations}}
As a widely used Physics assessment, various metrics and scores have been established for the FCME \cite{madsen2018}, but these are beyond the scope of the present study. Therefore, in terms of performance, only the percentage of correct answers is considered for each data set.\\
\\
For the data from the FMCE part of the instrument the following quantities were calculated:
\begin{enumerate}
	\item 	percentage of key (correct) answers for each question, $p_{K}$,
	\item 	percentage of different distractor (wrong) answers to each question, which is combined for each MCQ to calculate $dH'$ using Equation~\eqref{eq:3}
	\item 	using $p_{K}\ $and $dH'$ the $D$' and $\bar{D}$ are calculated from Equations~\eqref{eq:5} and \eqref{eq:6}.
\end{enumerate}
\hypertarget{evaluation-metric}{%
	\subsection{Evaluation metric}\label{evaluation-metric}}
Consistency of $\bar{D}$ between the two independently collected cohorts is assessed using Spearman's rank correlation coefficient, $\rho$, rather than a linear (Pearson) correlation \footnote{The data is expected to be non-linear and the Spearman is better suited to correlating direction and consistency.}. Spearman's $\rho$ ranks the $47$ FMCE items independently within each cohort by their $\bar{D}$ value and correlates the two rank orderings, so it tests whether the items identified as most concentrated or most random agree between cohorts, without requiring the raw $\bar{D}$ values themselves to match on the same numerical scale---the appropriate standard for a measure intended to rank items for triage rather than to reproduce an absolute score.
\hypertarget{results}{%
	\section{Results}\label{results}}
Applying the MCQ distractor analysis procedures to the FMCE component, both $\bar{D}$ and $p_{K}$ were calculated for each of the $47$ items. The label ``I'' refers to the 2021/2022 data and ``II'' to the 2022/2023 data, per Table~\ref{tab:2}. The results are shown in Figure~\ref{fig:4}, where each plot marker is labelled with the corresponding question number. As was done in Figure~\ref{fig:3}, Figure~\ref{fig:4} is divided into four quadrants.
\hypertarget{class-performance}{%
	\subsection{Class performance}\label{class-performance}}
Both classes performed very poorly in the assessment. The possible reasons for the poor performance, as well as consideration of the validity or applicability of the FCME to the course, are beyond the scope of this work.\\
\\
Based on the poor performance, most data points are thus clustered in Q2 and Q3 of Figure~\ref{fig:4}. Despite being two separate groups of students and sampled at very different rates (see Table~\ref{tab:2}) the data points in Q1 and Q4 are exactly the same for both data sets.
\hypertarget{degree-of-randomness}{%
	\subsection{Degree of randomness}\label{degree-of-randomness}}
For both I and II, data points in Figure~\ref{fig:4} concentrate in Q2 and Q3, consistent with the low scores captured in Table~\ref{tab:2}. Across all deployments, $\bar{D}$ values fall within a relatively narrow range: an upper bound of approximately $0.9$ and a lower bound of approximately $0.2$, with most data points lying between these two limits.
\subsection{Excluded items}
To connect the $\bar{D}$ values above to independent evidence about student cognition, each of the 47 FMCE items was cross-referenced against the FMCE question clusters as well as the named, interview-validated Newtonian misconceptions \cite{Smith2008}. That FMCE implementation guide \cite{madsen2018} also explicitly excludes seven items (5, 6, 15, 33, 35, 37, 39) from Newtonian-diagnostic scoring, for reasons unrelated to entropy or concentration---items 5, 6, and 15 are treated as reading-ability checks rather than physics-reasoning probes, and items 33, 35, and 37 are excluded from the Newton III cluster due to documented false-positive risk. This gives an independent, pre-existing classification of which items should carry a diagnosable misconception and which should not, against which the $\bar{D}$ values can be checked.
\begin{figure}[tbp]
	\centering
	\includegraphics[width=0.8\linewidth]{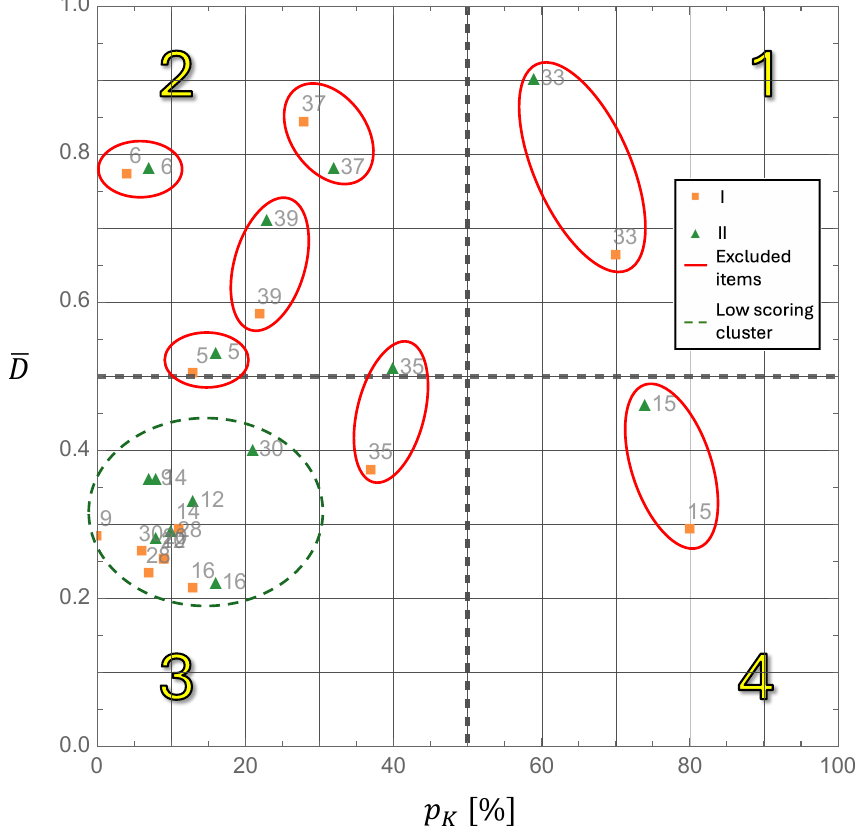}
	\caption{The excluded items as well as the low-scoring cluster of Table \ref{tab:excluded}.}
	\label{fig:5}
\end{figure}\\
\\
The position of the seven excluded items in Figure \ref{fig:4} are indicated in Figure \ref{fig:5} with the position of the item for both cohorts circled in red. Also indicated in Figure \ref{fig:5} is the cluster of low-scoring items ($p_{K}[\%]\lesssim20\%$ and $\bar{D}\leq0.4$) that repeats in both I and II, listed in Table \ref{tab:matchrates}.   Figure \ref{fig:5} two patterns emerge: First, all items in the low-scoring cluster are all members of a named clusters \cite{Smith2008}: 14, 16, and 20 (Force Graphs), 9, 12, and 28 (Reversing Direction), and 30 (Newton III), while none of the seven excluded items appear in this cluster.\\
\\
Second, the excluded items that have consistent low scoring $p_{K}[\%]$ values ($p_{K}[\%]\lesssim30\%$ for items 5, 6, 37, and 39) also show consistency in their $\bar{D}$ values across the cohorts, distinctly on the higher end in both cohorts relative to the low-scoring cluster (Table~\ref{tab:excluded}). These items were excluded from Newtonian scoring and question clusters \cite{madsen2018,Smith2008}, for reasons that have nothing to do with response entropy.
\begin{table}[htbp]
	\centering
	\caption{$\bar{D}$ for the four  low-scoring excluded items.}
	\label{tab:excluded}
	\begin{tabular}{ccccc}
		\toprule
		\multirow{2}*{ Item}&  $p_K[\%]$ & $\bar{D}$  & $p_K[\%]$ &  $\bar{D}$  \\
		&     (I)  &   (I)  &   (II)  &   (II) \\
		\midrule
		5  &13\%& 0.50 &16\%& 0.53 \\
		6  &4\%& 0.77 &7\%& 0.78 \\
		37 &28\%& 0.84 &32\%& 0.78 \\
		39 &22\%& 0.58 &23\%& 0.71 \\
		\bottomrule
	\end{tabular}
\end{table}\\
%\\
%
\subsection{Relating item-level randomness to documented misconceptions}
\subsubsection{Low-scoring cluster}
A further check is possible using the item-level response counts underlying the $\bar{D}$ values,
rather than the aggregate $\bar{D}$ and $p_K$ alone. Using \citet{Smith2008}'s
coded responses (their Tables IV---X) as related to Newtonian and non-Newtonian thinking, each item's actual distribution of wrong answers was checked against the specific answer option coded to that item's named non-Newtonian thinking misconception.
Across the cluster of low-scoring items highlighted above, a mean of 89\% (cohort I) and
88\% (cohort II) of \emph{all} incorrect responses landed on exactly the answer options coded to that
item's named misconception, replicated across two independently collected cohorts (Table~\ref{tab:matchrates}).
Across all 36 documented-cluster items with a defined responses, the mean match rate was 68\%
(cohort I) and 65\% (cohort II).
\begin{table}[htbp]
	\centering
	\caption{Percentage of incorrect responses matching the documented distractor option \cite{Smith2008}, for the seven-item low-scoring cluster ($p_{K}[\%]\lesssim20\%$ and $\bar{D}\leq0.40$ in both cohorts).}
	\label{tab:matchrates}
	\begin{tabular}{cccccc}
		\toprule
%		Item & Cluster &  $\bar{D}$(I)&$\bar{D}$(II)&Match \% (I) & Match \% (II) \\
		\multirow{2}*{ Item} &FMCE&  $\bar{D}$  &  $\bar{D}$   &  Match \%   &  Match \%  \\
		&  Cluster  &   (I)  &   (II)  &   (I)  &   (II) \\
		\midrule
		9  & Reversing Direction &0.28 &0.36 & 83 & 76 \\
		12 & Reversing Direction  & 0.25&0.33  & 86 & 80 \\
		14 & Force Graphs               & 0.29& 0.36 & 55 & 63 \\
		16 & Force Graphs            &0.21 &0.22  & 87 & 88 \\
		20 & Force Graphs            & 0.25&  0.28 & 82 & 82 \\
		28 & Reversing Direction  &0.23 &0.29 & 88 & 83 \\
		30 & Newton III               & 0.26& 0.40 & 89 & 77 \\
		\bottomrule
	\end{tabular}
\end{table}\\
\\
The rank ordering of $\bar{D}$ is stable between the two independently collected cohorts
(Spearman's $\rho = 0.78$ across all 47 items), and $\bar{D}$ is uncorrelated with $p_K$ within
each cohort ($r$ = $-$0.12 and $r = 0.10$ for cohorts I and II respectively).

\subsubsection{Ability-conditioned check}\label{subsec:split}
Using the underlying individual-student response data, each cohort's students were split at the median of their total score across all administered items, and the item-letter match rate and $\bar{D}$ were recomputed separately within each half, restricted in each case to students who answered that item incorrectly. For cohort I, the median score was 19\%, with split sample sizes depending on the question, with averages of $\bar{N}_L=25$ and $\bar{N}_H=22$.  For cohort II, median was 15\% and $\bar{N}_L=109$ and $\bar{N}_H=72$.
Table~\ref{tab:abilitysplit} reports the match rate for both halves of both cohorts, for the seven items in the low-scoring cluster.
\begin{table}[htbp]
	\centering
	\caption{Item-letter match rate (\%), split at the median of total score into lower (L)- and higher (H)-scoring halves.}
	\label{tab:abilitysplit}
	\begin{tabular}{c|c|c|c|c|c|c}
		\toprule
		\multirow{2}*{ Item} &  Match \%   &  Match \%   &  Match \%   &  Match \%  &  H$-$ L\%  &  H$-$ L\% \\
		 & (I, L) & (I, H) &  (II, L) & (II, H)& (I)&(II) \\
		\midrule
		9  & 80 & 86 & 74 & 76& 6& 2\\
		12 & 83 & 89 & 78 & 83 &7& 4\\
		14 & 25 & 82 & 56 & 76 &57& 20\\
		16 & 82 & 94 & 86 & 91 &12&5 \\
		20 & 75 & 89 & 82 & 84 &14& 2\\
		28 & 91 & 84 & 82 & 86&-7& 4\\
		30 & 83 & 95 & 74 & 80 &13& 6\\
		\bottomrule
	\end{tabular}
\end{table}
Of the seven items across two cohorts comparisons in Table~\ref{tab:abilitysplit}, the match rate is higher in the higher-scoring half in all but one; the one exception is item 28 in cohort I. The trend for $\bar{D}$ runs the opposite way, as expected, since randomly-selected answers are less likely among higher-performing students: Table~\ref{tab:splitDbar} shows $\bar{D}$ falling from the lower- to the higher-scoring half in 13 of the same 14 comparisons, with item 28 in cohort I again the sole exception---the same item-cohort pairing in both tables, and at a per-half sample size of $\bar{N}\approx20$--$25$ in cohort I where a single reversal is well within what chance alone could produce.
{
	
	\renewcommand{\arraystretch}{1.1}
	\setlength{\extrarowheight}{1pt}
\begin{table}[htbp]
	\centering
	\caption{Comparison split median, as in Table \ref{tab:abilitysplit}, $\bar{D}$ values}
	\begin{tabular}{c|c|c|c|c|c|c}
		\hline
		\multirow{2}*{ Item} & $\bar{D}$  &  $\bar{D}$  &H - L & $\bar{D}$ & $\bar{D}$ & H - L\\%[50pt]
		  & (I, L) & (I, H)  & (I)& (L, II)  &(H, II)  &(II)\\
		\hline
		9&0.31 & 0.23 & -0.08 & 0.38 & 0.36 & -0.02 \\
		12&0.26 & 0.22 & -0.04 & 0.34 & 0.30 & -0.04 \\
		14&0.40 & 0.20 & -0.21 & 0.37 & 0.28 & -0.09 \\
		16&0.23 & 0.16 & -0.07 & 0.24 & 0.19 & -0.04 \\
		20&0.36 & 0.18 & -0.18 & 0.31 & 0.26 & -0.06 \\
		28&0.20 & 0.24 &  0.04 & 0.30 & 0.27 & -0.03 \\
		30&0.32 & 0.20 & -0.12 & 0.43 & 0.36 & -0.07 \\
		\hline
	\end{tabular}
	\label{tab:splitDbar}
\end{table}

}

\subsubsection{Reversing Direction}
\begin{figure}[tbp]
\centering
\includegraphics[width=0.95\linewidth]{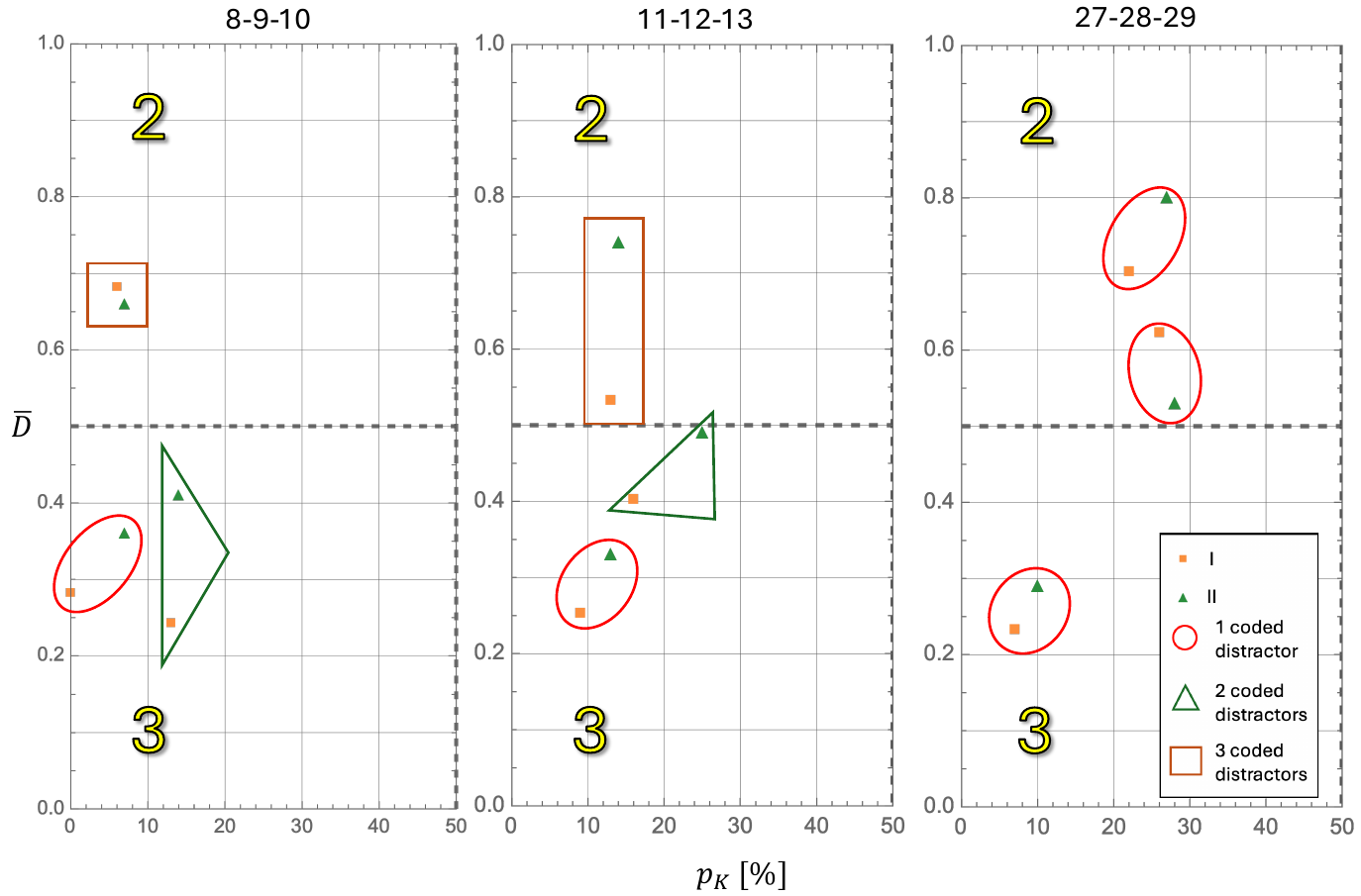}
\caption{$p_{K}[\%]$-$\bar{D}$ for the Reversing Direction triplets with the number of misconception coded distractors.}
\label{fig:6}
\end{figure}

The FMCE items 8--9--10, 11--12--13, and 27--28--29 each express a different stage of student reasoning about changes in direction of motion with reported common answer pattern associated with this misconception at each position of these triplets \cite{Smith2008, Thornton1996}. Of interest is the number of distractors coded to each misconception \footnote{A complication arises in applying this directly: the reported pattern characterises the general reasoning associated with each position, not the misconception in isolation, and hence also include the correct key for every item in the Reversing Direction cluster \cite{Smith2008}.}. Since $\bar{D}$ is computed only over incorrect responses, the key option is excluded from the count of common distractor responses at every position. The specific answer letters are not reproduced here, to avoid disclosing FMCE's answer key alongside its item numbers \cite{pportFCME2025}.\\
\\
The cases where misconception codes to multiple distractors, $\bar{D}$ should be larger than the case where the misconception codes to a single distractor since the expected incorrect student' responses should be more concentrated.  To investigate, Figure~\ref{fig:6} shows the $p_{K}[\%]$--$\bar{D}$ values for both cohorts on the Reversing Direction  triplets marked by this count: a circle for the single-distractor items, a triangle for the two-distractor items, and a rectangle for the three-distractor items. Table~\ref{tab:RD} reports $\bar{D}$ and the item-letter match rate for all nine items across the three triplets, together with the number of common distractor responses at that triplet position (Figure~\ref{fig:6}).
\begin{table}[htbp]
	\centering
	\caption{$\bar{D}$ and misconception match rate for the Reversing Direction triplets.}
	\label{tab:RD}
	\begin{tabular}{c|c|c|c|c}
		\toprule
		\multirow{2}*{ Item}  &  $\bar{D}$  &  $\bar{D}$   &  Match \%   &  Match \%  \\
		&      (I)  &   (II)  &   (I)  &   (II) \\
		\midrule
		8   & 0.68 & 0.66 & 80 & 80 \\
		9   & 0.28 & 0.36 & 83 & 76 \\
		10  & 0.24 & 0.41 & 98 & 86 \\
		11  & 0.53 & 0.74 & 73 & 70 \\
		12  & 0.25 & 0.33 & 86 & 80 \\
		13  & 0.40 & 0.49 & 76 & 78 \\
		27  & 0.71 & 0.80 & 75 & 67 \\
		28  & 0.23 & 0.29 & 88 & 83 \\
		29  & 0.62 & 0.53 & 59 & 67 \\
		\bottomrule
	\end{tabular}
\end{table}\\
\\
As a generalisation check the comparison is extended to two further FMCE item clusters: Force Sled cluster (items 1 - 4, 7) with only item 3 having multiple distractors coded to the same misconception and the Force Graphs cluster (items 14, 16 - 21) with items 17 and 21 coded to a broader model than the rest \cite{Smith2008}.  As reported in Table~\ref{tab:generalisation}, all items' $p_K<40\%$, including half of the previous identified low-scoring cluster (items 14, 16, and 20). Both the variation in the matched misconception distractor as well as the $\bar{D}$-values will be considered in the following Discussion section, specifically Section \ref{sample-agnostic-diagnostics}. 
\begin{table}[htbp]
	\centering
	\caption{$\bar{D}$ and item-letter match rate for the Force Sled and Force Graphs generalisation check.}
	\label{tab:generalisation}
	\begin{tabular}{c|c|c|c|c|c|c|c}
		\toprule
		\multirow{2}*{ Item} & \multirow{2}*{Cluster} & $p_K[\%]$ & $\bar{D}$ &   $p_K[\%]$ & $\bar{D}$  & Match \% & Match \%  \\
		 & &(I) & (I)  &  (II) & (II) & (I) & (II) \\
		\midrule
		1  & Force Sled    &38& 0.35 & 31 & 0.24 & 83 & 93 \\
		2  & Force Sled   &2 & 0.53 & 7 & 0.28 & 68 & 87 \\
		3  & Force Sled    &21& 0.65 & 19 & 0.62 & 68 & 77 \\
		4  & Force Sled    &27& 0.35 & 24 & 0.31 & 66 &86 \\
		7  & Force Sled    &34& 0.66 & 18 & 0.70 & 52 & 50 \\
		14 & Force Graphs &11 & 0.29 & 8 & 0.36 & 55 & 63 \\
		16 & Force Graphs  &13& 0.21 &  16& 0.22 & 87 & 88 \\
		17 & Force Graphs  &5& 0.54 & 9 & 0.62 & 45 & 47 \\
		18 & Force Graphs  &9& 0.66 & 9 & 0.51 & 14 & 2 \\
		19 & Force Graphs  &12& 0.56 &  13& 0.53 & 50 & 50 \\
		20 & Force Graphs  & 9&0.25   &8 & 0.28 & 82 & 82 \\
		21 & Force Graphs  &2& 0.73   & 9& 0.82 & 16 & 13 \\
		\bottomrule
	\end{tabular}
\end{table}
\hypertarget{discussion}{%
	\section{Discussion}\label{discussion}}
The application of dITA to a well-validated, widely used assessment in a low-stakes context revealed a consistent and striking trend: low-performing classes showed low-to-high randomness (values ranging from $0.2$ to $0.9$) in the distractor distributions. This pattern was sample-agnostic, appearing with similar strength across different participation rates ($12\%$ and $45\%$, as per Table \ref{tab:2}) and, most importantly, across two distinct student cohorts. It also matched the incorrect student responses in the categories of validated incorrect responses. This suggests a robust, context-driven behavior rather than a cohort-specific phenomenon anomaly, as is motivated by the following.
\hypertarget{positioning-ita-relative-to-instrument-validation-methods}{%
	\subsection{Positioning dITA relative to instrument-validation methods}\label{positioning-ita-relative-to-instrument-validation-methods}}
It is worth being explicit about what kind of tool dITA is, because the most directly comparable existing methods answer a related but distinct question. Concentration analysis \cite{bao2001} computes a structurally similar score-independent dispersion statistic (the concentration deviation, $\Gamma$) from distractor response counts, but does so as part of validating a research-grade instrument against misconceptions already catalogued through qualitative interview studies, typically pooling several hundred students across multiple classes. Full-information item analysis \cite{haladyna2021} goes further, conditioning distractor selection on total-score groups (trace lines) to separate genuine item flaws from student cognition. Both presuppose infrastructure---a validated instrument, an established misconception taxonomy, or a stable multi-class sample---that a lecturer analysing a single classroom quiz the day after it was administered does not have.\\
\\
dITA is offered for that gap specifically: a lightweight, single-administration triage step that helps an instructor decide, before any heavier analysis is undertaken, whether a class-wide wrong-answer pattern looks coherent enough to justify a targeted re-teach or diffuse enough to suggest a broader intervention. It is not proposed as a substitute for concentration analysis or full-information item analysis where those are feasible; rather, it is intended to flag which items are worth that closer, more resource-intensive scrutiny in the first place. Read this way, the entropy-based $\Gamma$-$\bar{D}$ comparison in the following sections is a difference in what each statistic is most sensitive to---$\Gamma$-type measures weight the size of the single largest wrong-answer cluster, which suits an instrument-validity question, while entropy weights how many distinct wrong answers are in play at all, which is closer to the pedagogical question of how targeted a re-teach can be--rather than a claim that one supersedes the other. Its role, in this sense, is to hand the instructor's own subject-matter expertise something concrete and inspectable to work from rather than to replace or outrank that expertise, and part of what makes that possible is that the method is legible enough to run in a spreadsheet.\\
\\
This same trajectory continues past concentration analysis itself, and in the same direction. \citet{bao2006} generalised concentration analysis into a fuller framework for representing and tracking the dynamics of student models, and \citet{wang2010} connected the same underlying question to item response theory directly--each extension adding modelling infrastructure rather than removing it. More recently, cognitive diagnostic modelling has taken this progression further still: a two-stage, hierarchy-extended MC-DINA model has been used to jointly estimate conceptual mastery and misconceptions from option-level multiple-choice responses on a forces-and-motion instrument \cite{jieDual2026}. Such models recover a considerably richer, per-student diagnostic profile than dITA claims to---but they require exactly the infrastructure dITA is designed to work without: model-fitting software, a validated item-to-construct mapping, and a sample well beyond a single classroom's. dITA is not offered as a competitor to this line of work, but as a much lighter tool for asking a related question, whether wrong answers reflect a shared misconception or diffuse confusion, at a much earlier moment.
\hypertarget{implications-for-measurement-theory}{%
	\subsection{Implications for Measurement Theory}\label{implications-for-measurement-theory}}
\hypertarget{a-new-signal-for-test-validity-context}{%
	\subsubsection{A New Signal for Test Validity \& Context}\label{a-new-signal-for-test-validity-context}}
For test developers, test validity in the context of IRT mainly concerns item analysis and distractor effectiveness \cite{Gierl2017}. The inclusion of random guessing in MixNRM-RG model \cite{jin2022} further enhances IRT models in this regard. Our results indicate that, in the particular low-stakes setting of the study, response patterns that appear statistically indistinguishable from random guessing can still be produced by students who are not genuinely engaged with the assessment. This means that standard psychometric models, which rely solely on response patterns, can misinterpret the underlying cognitive state driving those responses. dITA, by incorporating supplementary behavioral or engagement metrics, can help identify this specific issue of a lack of genuine cognitive engagement, whether brought on by the test and/or its administration context. The well-validated items are functioning, but the assessment context is not.
\hypertarget{sample-agnostic-diagnostics}{%
	\subsubsection{Sample-Agnostic Diagnostics}\label{sample-agnostic-diagnostics}}
The persistence of the low-to-moderate randomness pattern reported above, across two very different participation rates and student groups, highlights dITA's core strength. Unlike methods from CTT or IRT, which can be unreliable with small or atypical samples \cite{Demars2008}, dITA's approach provides a diagnosis that is a systemic characteristic of the assessment event. This allows an analysis that is not focused on which students attended or how many, but on how those who did attend approached the test.\\
\\
This is not only a qualitative impression. The item-level ranking of $\bar{D}$ values is substantially stable across the two independently collected cohorts (Spearman's $\rho = 0.78$ across all 47 items), and $\bar{D}$ is essentially uncorrelated with $p_K$ within each cohort ($r = -0.12$ and $r = 0.10$ for cohorts I and II respectively). Together, these indicate that $\bar{D}$ is picking up a stable property of the item itself, rather than sampling noise or a simple recoding of item difficulty.\\
\\
Cross-referencing against documented misconceptions \cite{Smith2008} corroborates this and shows where the stability comes from. Among the Table \ref{tab:matchrates} low-scoring cluster, a mean of 81\% (cohort I) and 79\% (cohort II) of incorrect responses matched the distractors coded to a specific misconception. The low-scoring excluded items, Table \ref{tab:excluded}, by contrast, show consistently higher $\bar{D}$ in both cohorts---consistent with their low $p_K$ reflecting general confusion with no clear choice of any answer, rather than a shared misconception. Thus reflecting the exact behaviour that precipitated their independent exclusion from Newtonian-diagnostic scoring. That a purely statistical measure recovers, unprompted, the same items an independent qualitative literature already flagged as diagnostically unreliable is a form of validation the entropy statistic was not designed to satisfy on its own.\\
\\
This correspondence also holds up under a crude check for ability confounding. Splitting each cohort's students at the median of their total score and recomputing the match rate separately within each half (Table~\ref{tab:abilitysplit}) shows the same direction in 13 of 14 comparisons: among students who answered a low-scoring-cluster item incorrectly, those with a higher overall score were more likely, not less, to have chosen the specific documented misconception letter, with $\bar{D}$ correspondingly lower in the higher-scoring half. This is the pattern a misconception account predicts, and an ability-independent artefact---an unusually attractive distractor pulling in students regardless of competence---does not: if the latter were driving the low $\bar{D}$ values, the match rate would be flat across the split, or higher among lower-scoring students more susceptible to a merely plausible-looking option. Instead, the signal concentrates specifically among students with more, not less, command of the material, consistent with a coherent, if incomplete, model rather than an item-design artefact or guessing. This is not a substitute for full trace-line conditioning across the whole ability range---the split here is a single median cut within an already low-scoring population, at modest per-half sample sizes in cohort I---but it is a direct test of, and a direct answer to, the one alternative explanation the aggregate misconception to distractor/s match-rate result on its own cannot rule out.\\
\\
A further, more granular illustration of this same sample-independence comes from the Reversing Direction cluster's own internal structure. Individual misconception are associated with a different number of common distractor responses at each position of the 8--9--10 and 11--12--13 triplets \cite{Smith2008}, the same underlying misconception is predicted to register different $\bar{D}$ values purely as a function of item design, independent of how coherently students actually reason. Figure~\ref{fig:6} shows that this predicted grouping separates cleanly as a function of $\bar{D}$ in both cohorts (representing different sampling rates)---a second, mechanistically grounded instance of the sample-agnostic behaviour above, independent of, and more granular than, the aggregate rank-correlation result.\\
\\
This same triplet structure is also a caution against reading $\bar{D}$ in isolation. Table~\ref{tab:RD} shows that items with markedly lower $\bar{D}$ than their neighbours within a triplet do not show a correspondingly higher item-letter match rate: match rates are broadly comparable within each triplet, despite a more than twofold spread in $\bar{D}$. The gap in $\bar{D}$ is driven almost entirely by how many distractor letters the single misconception happens to occupy at each triplet position, not by a difference in how unified student thinking actually is. Read without the crosswalk to \citet{Smith2008}'s misconception to distractor coding, $\bar{D}$ alone would rank these items' conceptual coherence incorrectly.\\
\\
Checking whether this effect generalises beyond the Reversing Direction cluster shows that it does not, and shows why that matters. Table~\ref{tab:generalisation} shows that in the Force Sled cluster, the item coded to the narrower model does not have the lower $\bar{D}$; its own match rate is simply lower, and this dominates instead. In the Force Graphs cluster, the two broader-model items diverge sharply from one another despite carrying the same model breadth, and the exceptionally low match rate on item 18---the lowest observed anywhere in this analysis---corresponds to its high $\bar{D}$, suggesting \citet{Smith2008}'s misconception to distractor coding for this item may simply not describe this population's dominant error. Taken together, these checks show that a low $\bar{D}$ consistently and specifically tracks alignment with a documented misconception, while a comparatively high $\bar{D}$ is a less specific signal that can arise for more than one reason---item design, poor fit to the documented model, or genuine guessing among them. This is, if anything, reassuring for $\bar{D}$'s validity as a diagnostic measure: it is precisely in the low-$\bar{D}$ region, where the paper's diagnostic claims are concentrated, that the correspondence with independently documented misconceptions holds most reliably.
\hypertarget{implications-for-measurement-practice}{%
	\subsection{Implications for Measurement Practice}\label{implications-for-measurement-practice}}
For decades, educational measurement has regarded the multiple-choice test as a straightforward tool: a student's score was considered a direct reflection of their knowledge \cite{Crocker2008}. Psychometricians crafted sophisticated models to refine this measurement, accounting for noise from random guessing and imperfections in test items \cite{abu2023,briggs2006,Gierl2017}. The aim was always to isolate the true signal of ability from the surrounding noise. However, a persistent, frustrating problem remained for anyone teaching: a poor test result yielded a number but no clear explanation to the most vital question---why? Did students fail because they held deep, stubborn misconceptions, or were they simply lost, guessing in the dark?\\
\\
This diagnostic silence has now possibly been broken. By applying the lens of Information Theory to student answer patterns, it can be measured not just whether students are wrong, but how they are wrong. The key metric is entropy---in this instance, used to quantify how a class's incorrect answers are distributed across the available distractors, not to identify what any of those answers mean. Low $\bar{D}$ is accordingly a flag, not a detector: it marks an item where incorrect answers concentrated on few options, which is worth checking against an independent source (a documented misconception, or the analyst's own subject-matter judgement) before concluding anything further. High $\bar{D}$ is the weaker signal of the two---it is consistent with unstructured guessing, but, as Section~\ref{sample-agnostic-diagnostics} shows directly, a genuine, well-documented misconception can also register as a high $\bar{D}$ if it happens to be expressed across several distractor options rather than one, so a high value should not by itself be read as ruling a misconception out. This simple, sample-independent calculation offers instructors a starting point for that closer look: is the problem, on the face of it, one of coherence or chaos?\\
\\
For educational practitioners, this acts as a wake-up call to revise their approach. The raw score no longer offers sufficient insight. The new focus is on diagnosing before prescribing. A low score combined with high, random error levels signals a critical warning for engagement issues, not content comprehension. It necessitates interventions related to motivation, test framing, and relevance before any re-teaching takes place. For test developers and institutions, it broadens the very definition of validity. A test must now demonstrate not only that it measures the construct but also that it can reliably provoke engaged responses from students in its intended context. These implications hold only once, as is discussed next, item-writing flaws and ability-related explanations have been ruled out; dITA identifies which items warrant that closer inspection rather than replacing it.
\hypertarget{possible-shortcomings}{%
	\subsection{Possible shortcomings}\label{possible-shortcomings}}
$\bar{D}$ illustrates concentrations in the incorrect answer states but cannot identify the origin of these concentrations. Therefore, concentration on incorrect answers should not automatically be attributed to a misconception about the question itself, as it may also result from other factors, some of which will be briefly discussed.\\
\\
A related concern is specific to linked sub-questions, where a later question's correct answer depends on the answer to, or a calculation from, an earlier one. $\bar{D}$ measures the pattern of incorrect responses to a single question in isolation; it has no way to see whether that pattern originates in the question itself or is inherited from an error made upstream. If the same underlying error propagates through a sequence of dependent sub-questions---an incorrect intermediate value carried into a dependent calculation, for instance---the same wrong answer will recur at each dependent step. This produces a concentrated, low $\bar{D}$ pattern that reflects one error appearing repeatedly, not several students independently converging on a shared misconception at that specific step, and it requires no misreading of the sub-question itself: a single upstream error is sufficient. Where sub-questions are linked in this way, a low $\bar{D}$ on a later item should be checked against the pattern on the question(s) it depends on before being read as an independent misconception signal. Interpreting the cause of a low-$\bar{D}$ pattern in linked sub-questions accordingly depends on the analyst's expertise and familiarity with the subject matter, not on $\bar{D}$ alone.\\
\\
A related concern is item quality itself. A concentrated wrong-answer pattern is also produced by item-writing flaws that have nothing to do with student cognition: a miskeyed correct response, an unintentionally plausible second correct option, or an implausible, never-selected distractor. As above, dITA does not itself adjudicate distractor validity: that remains the province of established IRT and CTT analysis, not a job for $\bar{D}$ here. dITA cannot, by itself, distinguish a genuine shared misconception from a broken item; a low $\bar{D}$ value should be read as a prompt to inspect the item, including with those established methods, before concluding a misconception is present.\\
\\
A further concern is that $\bar{D}$ is computed from the pooled class response without conditioning on individual students' overall ability, so a concentrated wrong-answer pattern could in principle reflect an unusually attractive distractor pulling in students across the ability range, rather than a genuine misconception concentrated among students who do not know the material. Where individual-student response data were available, we tested this directly: splitting each cohort at the median of total score and recomputing the item-letter match rate within each half, as discussed in Section \ref{subsec:split}, shows the match rate consistently higher, and $\bar{D}$ correspondingly lower, among the higher-scoring half---the opposite of what an ability-independent distractor artefact would predict. This is informative but partial evidence, not a substitute for full trace-line conditioning across the whole ability range: the check available here is a single median split within an already low-scoring population, at modest per-half sample sizes in the smaller cohort, and it was only possible because individual-student data happened to be available for this deployment. Extending $\bar{D}$ to condition on total-score groups more finely, in the spirit of full-information item analysis, remains the most direct way to close this gap in general, and is accordingly the most important item on the future-work agenda in Section~\ref{future-work}.
\hypertarget{conclusions}{%
	\section{Conclusions}\label{conclusions}}
This paper set out to answer the question: \emph{How does an information-theoretic measure of uncertainty in distractor choice deliver fast, practical, and robust diagnostic feedback for MCQ-based instruction---overcoming a core weakness of conventional psychometrics in classroom settings?} The proposed solution was a novel distractor analysis method that compares the entropy of a question's actual distractor distribution to that of a random distribution.\\
\\
This method effectively shifts the analytical focus to the group level, measuring the emergent degree of randomness (aleatory uncertainty) in the class's incorrect answers. It enables instructors to diagnostically assess class performance, flagging questions where incorrect answers concentrate on a specific option for closer inspection, as distinct from those where poor results stem from widespread guessing.\\
\\
A key advantage of this method is that it is not restricted by sample size or statistical significance, overcoming a major limitation of traditional psychometric techniques. As a result, it provides a powerful and practical tool that allows educators to go beyond merely identifying difficult questions to understanding the underlying reasons for student performance. This enables more targeted and effective pedagogical interventions. In doing so, it bridges the gap between advanced measurement theory and real-world classroom practices, clearly indicating whether issues stem from the curriculum or from the conditions under which it is assessed. This interpretation is not merely asserted: cross-referencing item-level response data against independently documented misconceptions on the FMCE shows that a majority of incorrect responses on the most concentrated items select the exact response coded to that item's named, interview-validated misconception, while items independently flagged elsewhere as diagnostically unreliable are exactly the items the measure identifies as having the highest degree of randomness.\\
\\
The contribution of dITA is therefore best understood as complementary to, rather than competing with, established distractor- and ability-conditioned analyses such as concentration analysis \cite{bao2001} or full-information item analysis \cite{haladyna2021}. Where those methods presuppose a validated instrument, a stable multi-class sample, or a pre-existing catalogue of misconceptions, dITA is designed to run on a single, ad hoc classroom administration and to flag which items justify that heavier, confirmatory investigation.
\hypertarget{future-work}{%
	\subsection{Future work}\label{future-work}}
Future work should extend dITA to condition entropy estimates on total-score groups (in the spirit of item trace lines), apply a finite-sample entropy correction for small classes, and evaluate the method on well-instructed, high-performing items, so that its diagnostic claims can continue to be checked against ground truth rather than against the entropy statistic alone.  This work will be done in the form of expanding the released computational tool \cite{DienerGitHub} as well as the mentioned manuscript in preparation regarding $\bar{D}$ by its corresponding Hill number.
\section{Acknowledgement}
The collection method and handling of the collected data were conducted in accordance with procedures approved by the Stellenbosch University Research Ethics Committee (REC): Social, Behavioural and Education Research (REC: SBE), as Project number 22314.
\section{Bibliography}\label{bibliography}
	\bibliographystyle{apsrev4-2}
\bibliography{PRPER}
\end{document}